\documentclass[12pt]{article}
\usepackage{epsfig}
\usepackage{hyperref}
\usepackage{amssymb}
\usepackage{amsmath}
\usepackage{amsfonts}
\usepackage{latexsym}
\usepackage{wasysym}
\usepackage{multirow}
\usepackage{fixmath}
\usepackage{color}
\usepackage{stackrel}
\usepackage{txfonts}                                                            
\usepackage{bbm}
\newcommand{\bea}{\begin{eqnarray}}
\newcommand{\eea}{\end{eqnarray}}
\newcommand{\bite}{\begin{itemize}}
\newcommand{\eite}{\end{itemize}}

\begin{document}

\title{
\vspace{-2.5cm} 
\flushleft{\normalsize ADP-15-6/T908} \\
\vspace{-0.35cm}
{\normalsize DESY 15-018} \\
\vspace{-0.35cm}
{\normalsize Edinburgh 2015/01} \\
\vspace{-0.35cm}
{\normalsize Liverpool LTH 1034} \\
%\vspace{-0.35cm}
%{\normalsize February 2015} \\
\vspace{0.75cm}
\centering{\Large \bf The electric dipole moment of the neutron from $\mathbf{2+1}$ flavor lattice QCD}\\} 
\author{F.-K.~Guo$^a$, R.~Horsley$^b$, U.-G.~Mei{\ss}ner$^{a,c}$, Y.~Nakamura$^d$, H.~Perlt$^e$,\\
P.~E.~L.~Rakow$^f$, G.~Schierholz$^g$, A.~Schiller$^e$ and J.~M.~Zanotti$^h$\\[1em] 
$^a$ Helmholtz Institut f\"ur Strahlen- und Kernphysik and Bethe Center\\ for Theoretical Physics, Universit\"at
Bonn, 53115 Bonn, Germany\\ [0.15em]
$^b$ School of Physics and Astronomy, University of Edinburgh,\\ Edinburgh
EH9 3FD, United Kingdom\\[0.15em] 
$^c$ Institute for Advanced Simulation, Institut f\"ur Kernphysik and J\"ulich\\ Center for Hadron Physics, 
JARA-FAME and JARA-HPC,\\ Forschungszentrum J\"ulich,
52425 J\"ulich, Germany\\[0.15em]
$^d$ RIKEN Advanced Institute for Computational Science,\\ Kobe, Hyogo 650-0047, Japan\\[0.15em] 
$^e$ Institut f\"ur Theoretische Physik, Universit\"at Leipzig,\\ 04103
Leipzig, Germany\\[0.15em]  
$^f$ Theoretical Physics Division, Department of Mathematical Sciences,\\
University of Liverpool, Liverpool L69 3BX, United Kingdom\\[0.15em] 
$^g$ Deutsches Elektronen-Synchrotron DESY,\\ 22603 Hamburg, Germany\\[0.15em] 
$^h$ CSSM, Department of Physics, University of Adelaide,\\ Adelaide SA 5005, Australia}

\date{}

\maketitle

%\begin{center}
%{\large QCDSF Collaboration}
%\end{center}

%\clearpage
\begin{abstract}
We compute the electric dipole moment $d_n$ of the neutron from a fully dynamical simulation of lattice QCD with $2+1$ flavors of clover fermions and nonvanishing $\theta$ term. The latter is rotated into a pseudoscalar density in the fermionic action using the axial anomaly. To make the action real, the vacuum angle $\theta$ is taken to be purely imaginary. The physical value of $d_n$ is obtained by analytic continuation. We find $\displaystyle d_n = -3.9(2)(9) \, \times 10^{-16} \, \theta \,e \,\mbox{cm}$, which, when combined with the experimental limit on $d_n$, leads to the upper bound $|\theta| \lesssim 7.4 \times 10^{-11}$.
\end{abstract}

\clearpage
\section{Introduction}

The electric dipole moment  $d_n$ of the neutron provides a unique and sensitive probe to physics beyond the Standard Model. It has played an important part over many decades in shaping and constraining numerous models of CP violation.
While the CP violation observed in $K$ and $B$ meson decays can be accounted for by the phase of the CKM matrix, the baryon asymmetry of the universe cannot be described by this phase alone, suggesting that there are additional sources of $CP$ violation awaiting discovery. 

\begin{figure}[b!]
\vspace*{-1.0cm}
\begin{center}
\epsfig{file=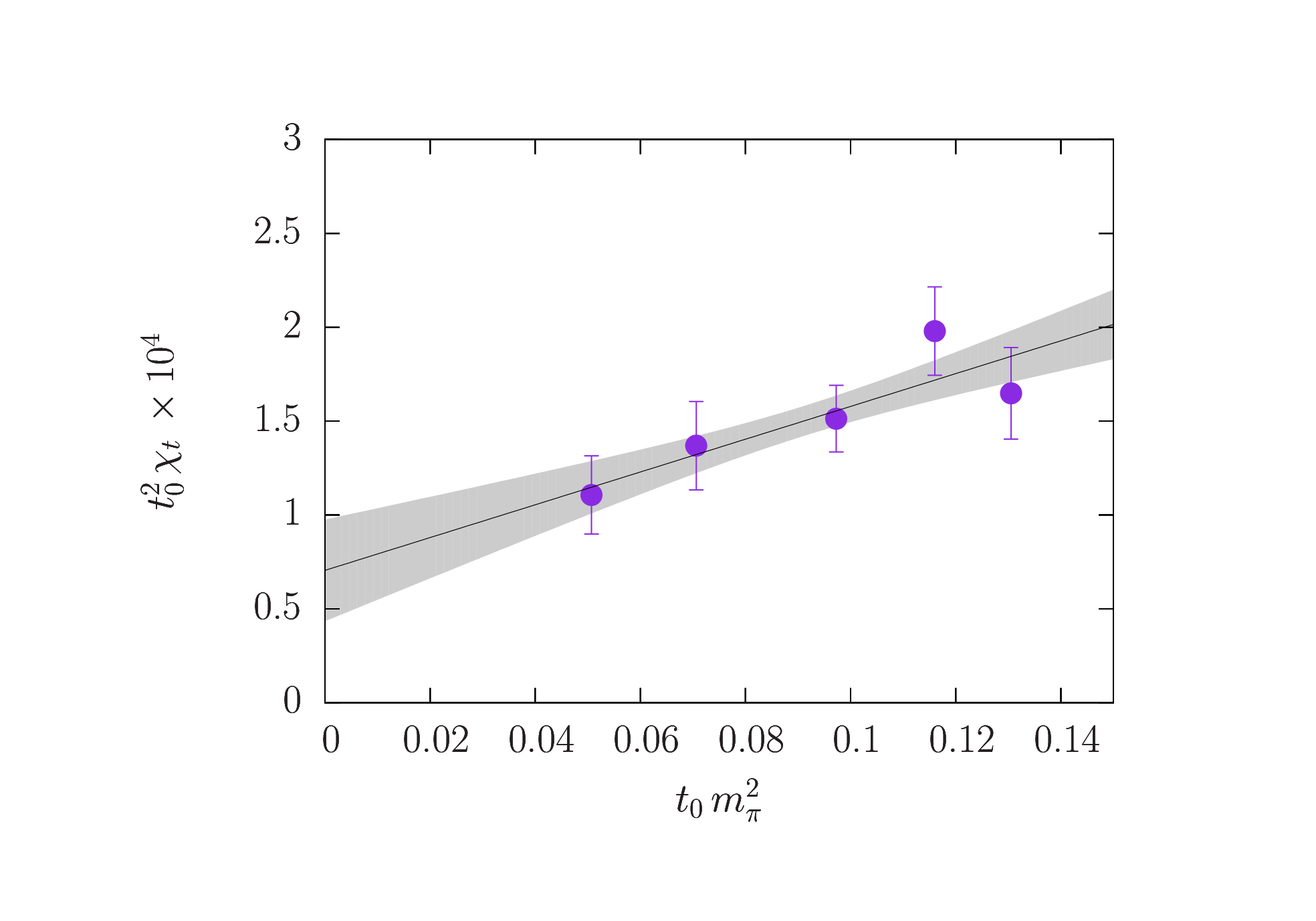,width=13cm,clip=}
\end{center}
\vspace*{-1.25cm}
\caption{The topological susceptibility on the SU(3) symmetric line $m_u=m_d=m_s$ as a function of $m_\pi^2$ in units of $t_0$.}
\label{fig1}
\end{figure}

QCD allows for CP-violating effects that propagate into the hadronic sector via the so-called $\theta$ term $S_{\theta\,}$ in the action,   
\begin{equation}
S=S_0+S_\theta \,, \quad S_\theta = i\,\theta\,Q \,,
\label{Stheta}
\end{equation}
where (in lattice notation)
\begin{equation}
Q=-\frac{1}{64\pi^2}\, \epsilon_{\mu\nu\rho\sigma}\, a^4 \sum_x F_{\mu\nu}^a F_{\rho\sigma}^a \in \mathbb{Z}
\end{equation} 
is the topological charge, and $S_0$ is the standard CP-preserving QCD action. Thus, there is the possibility of strong $CP$ violation arising from a nonvanishing vacuum angle $\theta$. In a wide class of GUTs the diagrams that generate a high baryon to photon asymmetry contribute to the renormalization of $\theta$, and hence to the electric dipole moment of the neutron. With the increasingly precise experimental efforts to observe the electric dipole moment~\cite{Harris,Lamoreaux:2009zz,Hewett:2012ns}, it is important to have a rigorous calculation directly from QCD.

It is practically impossible to perform Monte Carlo simulations with the action (\ref{Stheta}) in four dimensions for any sensible definition of the topological charge and any angle $|\theta|>0$. Absorbing the $\theta$ term into the observable~\cite{Berruto:2005hg,Shintani:2008nt} is not a viable alternative, as $\langle Q^2\rangle$ is found not to vanish if one of the quark masses is taken to zero at present values of the coupling. In Fig.~\ref{fig1} we show the topological susceptibility $\chi_t=\langle Q^2\rangle/V$ on $32^3\times 64$ lattices taken from \cite{Bietenholz:2011qq} at spacing $a=0.074\,\mbox{fm}$. The charge $Q$ has been computed from the Wilson flow~\cite{Luscher:2010iy} at flow time $t_0$. Similar results have been reported in~\cite{Bruno:2014ova}. As a result, $d_n$ will not vanish in the limit of zero quark mass either, except perhaps for chiral fermions. Exactly that was found in~\cite{Shintani:2012uba}. This precludes a meaningful extrapolation of $d_n$ to the physical point. There are indications that the situation will improve for lattice spacings $a \lesssim 0.04\,\mbox{fm}$ only~\cite{Bruno:2014ova}. 

It so happens that the $\theta$ term can be chirally rotated into the fermionic part of the action, making use of the axial anomaly~\cite{Baluni:1978rf}. The outcome of that is 
\begin{equation}
S_\theta=-\frac{i}{3}\,\theta\; \hat{m}\; a^4 \sum_x \left(\bar{u}\gamma_5 u + \bar{d}\gamma_5 d + \bar{s}\gamma_5 s\right)\,, \quad \hat{m}^{-1} = \frac{1}{3} \left(m_u^{-1}+m_d^{-1}+m_s^{-1}\right)
\label{Ftheta}
\end{equation} 
%where
%\begin{equation}
%\hat{m}^{-1} = \left(m_u^{-1}+m_d^{-1}+m_s^{-1}\right)\,.
%\end{equation}
for three quark flavors with nondegenerate masses. This action lends itself to numerical simulations for imaginary values of $\theta$~\cite{Horsley:2008gv}. As we are mainly interested in small values of $\theta$, the results can be analytically continued to real numbers without difficulties, assuming that the theory is analytic in the vicinity of $\theta=0$. 

In this paper we present an entirely dynamical calculation of the electric dipole moment of the neutron  on the lattice. This is a challenging task. As $d_n$ quickly diminishes towards physical quark masses, the angle $\theta$ has to be chosen increasingly larger to compensate for that. This in turn leads to a substantial increase of zero modes, which slows down the simulations substantially and eventually will result in exceptional configurations~\cite{Schierholz:1998bq}.

\section{The simulation}

We follow~\cite{Bietenholz:2010jr,Bietenholz:2011qq} and start from the SU(3) flavor symmetric point $m_u=m_d=m_s\equiv m_0$, where $m_\pi=m_K$. Our strategy has been to keep the singlet quark mass $\bar{m} = (m_u+m_d+m_s)/3$ fixed at its physical value, while $\delta m_q=m_q-\bar{m}$ is varied. As we move from the symmetric point to the physical point along the path $\bar{m}=\mbox{constant}$, the $s$ quark becomes heavier, while the $u$ and $d$ quarks become lighter. These two effects tend to cancel in any flavor singlet quantity, such as the topological susceptibility $\chi_t=\langle Q^2\rangle/V$. The cancellation is perfect at the symmetric point~\cite{Bietenholz:2011qq}. 

We assume $u$ and $d$ quarks to be mass degenerate, writing $m_\ell = m_u=m_d$. The vacuum angle is taken purely imaginary,
\begin{equation}
\theta = i\,\bar{\theta} \,.
\end{equation}
This leads us to consider the action 
\begin{equation}
S_\theta=\bar{\theta}\; \frac{m_\ell\, m_s}{2m_s+m_\ell} \, a^4 \sum_x \left(\bar{u}\gamma_5 u + \bar{d}\gamma_5 d + \bar{s}\gamma_5 s\right)\,,
\label{Itheta}
\end{equation}
which is real and vanishes at $m_\ell=0$ as well as $m_s=0$.

Our fermion action has single level stout smearing for the hopping terms together with unsmeared links for the clover term. With the (tree level) Symanzik improved gluon action this constitutes the Stout Link Non-perturbative Clover or SLiNC action~\cite{Cundy:2009yy}. To cancel O(a) terms the clover coefficient $c_{SW}$ has been computed nonperturbatively. For each flavor the fermion action to be simulated reads
\begin{equation}
S^q = S_0^q + S_\theta^q = a^4 \sum_x \bar{q}\left(D-\frac{1}{4} c_{SW}\, \sigma_{\mu\nu}\,F_{\mu\nu} + m_q + \frac{\lambda}{2a}\,\gamma_5\right)q \,,
\label{Saction}
\end{equation}
where $D$ is the Wilson Dirac operator and 
\begin{equation}
\lambda = \bar{\theta}\, 2a \, \frac{m_\ell\, m_s}{2m_s+m_\ell} \,.
\label{lambda}
\end{equation}
The extra term in the action (\ref{Saction}) can be treated in a similar way as we treat disconnected diagrams in calculations of singlet hadron matrix elements and renormalization factors~\cite{Chambers:2014qaa,Chambers:2014pea}.
We use BQCD~\cite{Nakamura:2010qh} to update the gauge fields. The calculations are done on $24^3\times 48$ lattices at $\beta=5.50$. At this coupling the lattice spacing was found to be $a=0.074(2)\,\mbox{fm}$~\cite{Horsley:2013wqa}, using the center of mass of the nucleon octet to set the scale. The parameters of the simulations are listed in Table~\ref{tab1}. Each ensemble consists of $O(2000)$ trajectories. The quark masses on the $\bar{m}=\mbox{constant}$ line are given by $m_q=1/2\kappa_q-1/2\kappa_{0,c}$ with $\kappa_{0,c}=0.12110$~\cite{Bietenholz:2011qq}.

\begin{table}[t]
\begin{center}
\begin{tabular}{c|c|c|c|c|c|c}
$\#$ &$\kappa_\ell$ & $\kappa_s$ & $am_\pi$ & $am_K$ & $am_N$ & $\lambda$ \\ \hline
1 & 0.12090 & 0.12090 & 0.1747(5) & 0.1747(5) & 0.4673(27) & 0.003 \\
2 & 0.12090 & 0.12090 & 0.1747(5) & 0.1747(5) & 0.4673(27) & 0.005 \\
3 & 0.12104 & 0.12062 & 0.1349(5) & 0.1897(4) & 0.4267(50) & 0.003 \\
4 & 0.12104 & 0.12062 & 0.1349(5) & 0.1897(4) & 0.4267(50) & 0.005 
\end{tabular}
\end{center}
\caption{The simulation parameters with $\bar{m}=\mbox{constant}$. The hadron masses refer to $\lambda=0$.}
\label{tab1}
\end{table}

\begin{figure}[t!]
\vspace*{-1.25cm}
\begin{center}
\epsfig{file=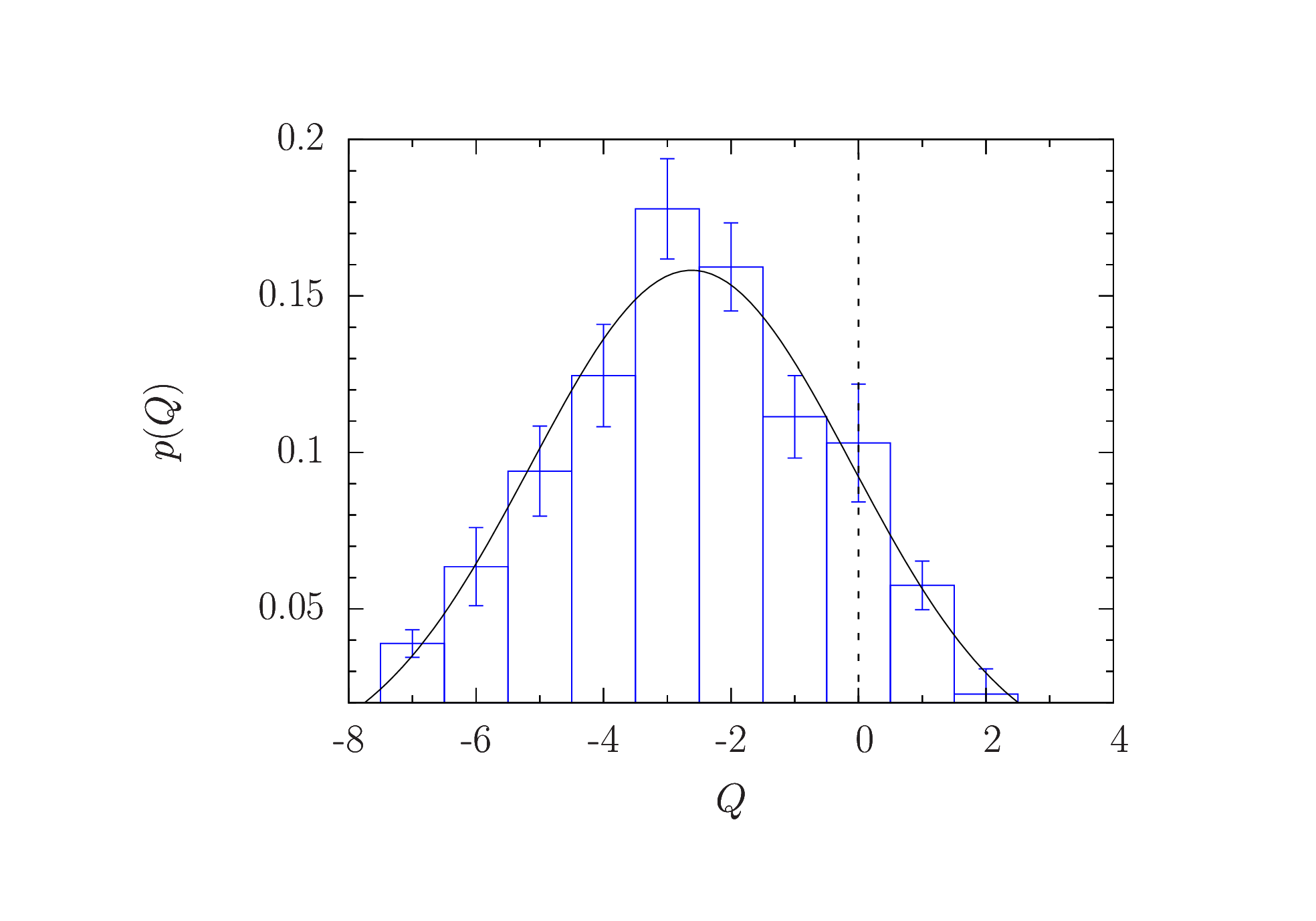,width=13cm,clip=}
\end{center}
\vspace*{-1cm}
\caption{The topological charge distribution $p(Q)$ (with $\sum_Q p(Q)=1$) of ensemble 4 at $\kappa_\ell=0.12104$, $\kappa_s=0.12062$ and $\lambda=0.005$, together with a Gaussian fit.}
\label{fig_hist}
\end{figure}

\begin{figure}[t!]
\vspace*{-1.25cm}
\begin{center}
\epsfig{file=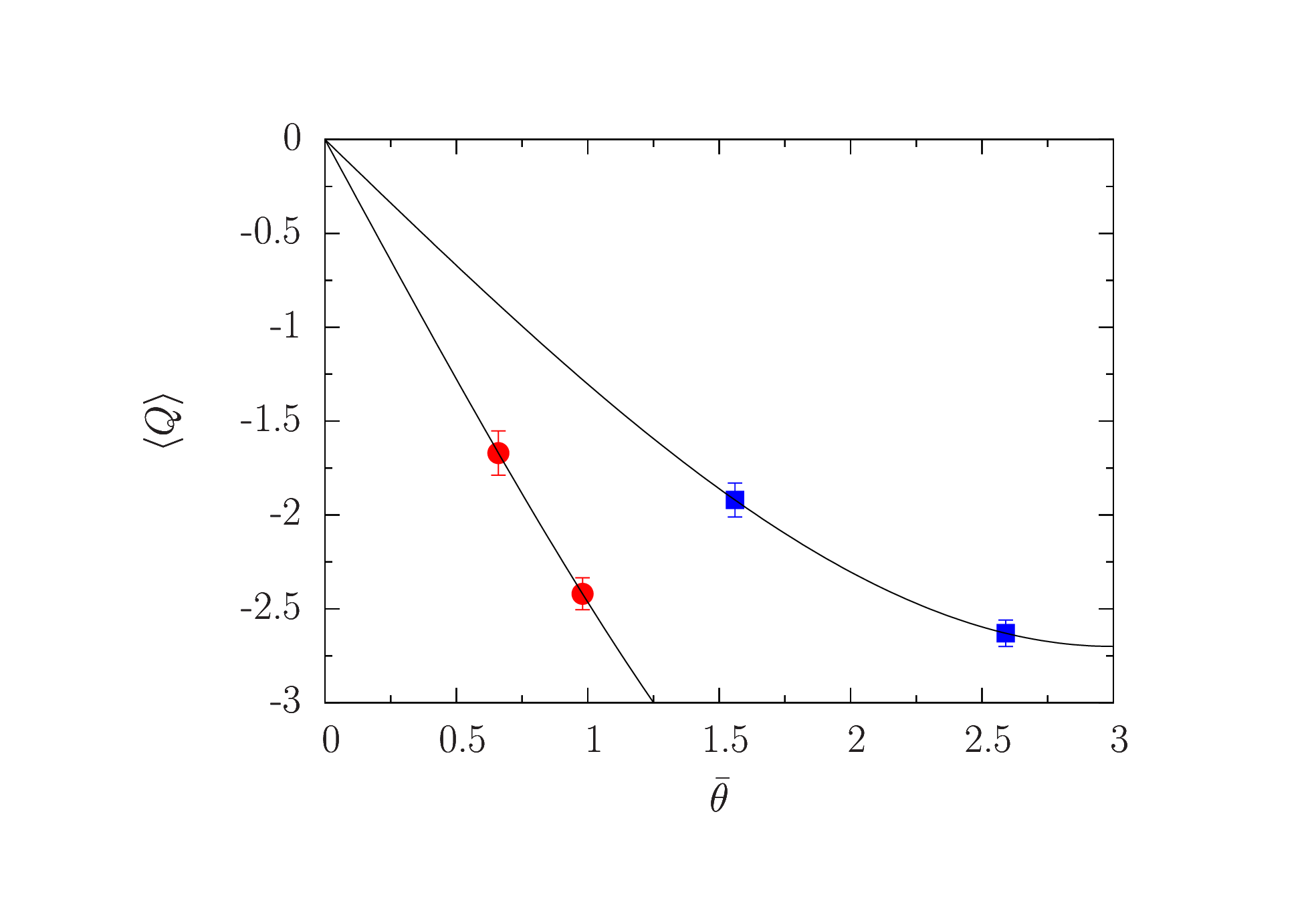,width=13cm,clip=}
\end{center}
\vspace*{-1cm}
\caption{The average charge $\langle Q\rangle$ as a function of $\theta$ for ensembles 1 and 2 (\textcolor{red}{$\CIRCLE$}) and ensembles 3 and 4 (\textcolor{blue}{$\blacksquare$}), together with linear plus cubic fits.}
\label{fig_Qtheta}
\end{figure}

We expect our ensembles to carry nonvanishing topological charge, $\langle Q\rangle \propto - \,\bar{\theta} \,\langle Q^2\rangle_c$, with $\langle Q^2\rangle_c = \langle Q^2\rangle - \langle Q\rangle^2 \propto \hat{m}$~\cite{Leutwyler:1992yt}. In Fig.~\ref{fig_hist} we show the charge histogram for ensemble 4, together with a Gaussian fit. As before, the topological charge has been computed from the Wilson flow at flow time $t_0$~\cite{Luscher:2010iy}. Evidently, $Q$ peaks at negative values. In Fig.~\ref{fig_Qtheta} we show $\langle Q\rangle$ as a function of $\bar{\theta}$ for both sets of quark masses, together with linear plus cubic fits. We find the slopes of the individual curves to be approximately proportional to $\hat{m}$, as expected.

\section{The evaluation}

At nonvanishing vacuum angle $\theta$ the nucleon matrix element of the electromagnetic current reads in Euclidean space
\begin{equation}
\langle p^\prime,s^\prime|J_\mu|p,s\rangle =
\bar{u}_\theta(\vec{p}^{\,\prime},s^\prime)\, \mathcal{J}_\mu\,
u_\theta(\vec{p},s)\,,
\label{me}
\end{equation}
where
\begin{equation}
\mathcal{J}_\mu = \gamma_\mu F_1^\theta(q^2) + \sigma_{\mu\nu} q_\nu
\frac{F_2^\theta(q^2)}{2m_N^\theta} +  (\gamma q\, q_\mu -
  \gamma_\mu \, q^2) \, \gamma_5 \, F_A^\theta(q^2) + \sigma_{\mu\nu} q_\nu \,
  \gamma_5 \frac{F_3^\theta(q^2)}{2m_N^\theta}
\label{current}
\end{equation}
and $q=p^\prime - p$, $q^2 = (\vec{p}^{\,\prime} -
\vec{p})^2 - (E^{\theta\,\prime} - E^\theta)^2$. In the $\theta$ vacuum the Dirac spinors pick up a phase~\cite{Shintani:2005xg},
\begin{equation}
\begin{split}
u_\theta(\vec{p},s) &= {\rm e}^{i\alpha(\theta)\gamma_5}\, u(\vec{p},s)\,,\\
\bar{u}_\theta(\vec{p},s) &= \bar{u}(\vec{p},s)\, {\rm
  e}^{i\alpha(\theta)\gamma_5}\,, 
\end{split}
\end{equation}

\clearpage

\clearpage
\noindent
so that 
\begin{equation}
\sum_s u_\theta(\vec{p},s)\, \bar{u}_\theta(\vec{p},s) = {\rm
  e}^{i\alpha(\theta)\gamma_5} \left(\frac{-i \gamma p + m_N^\theta}{2
  E_N^\theta}\right) {\rm  e}^{i\alpha(\theta)\gamma_5}
\end{equation}
with $\gamma p = \vec{\gamma}\vec{p} + i E \gamma_4$. The electric dipole moment
is given by
\begin{equation}
d_n = \frac{e\,F_3^\theta(0)}{2m_N^\theta} \,.
\label{edm}
\end{equation}

\begin{figure}[t!]
\vspace*{-0.5cm}
\begin{center}
\epsfig{file=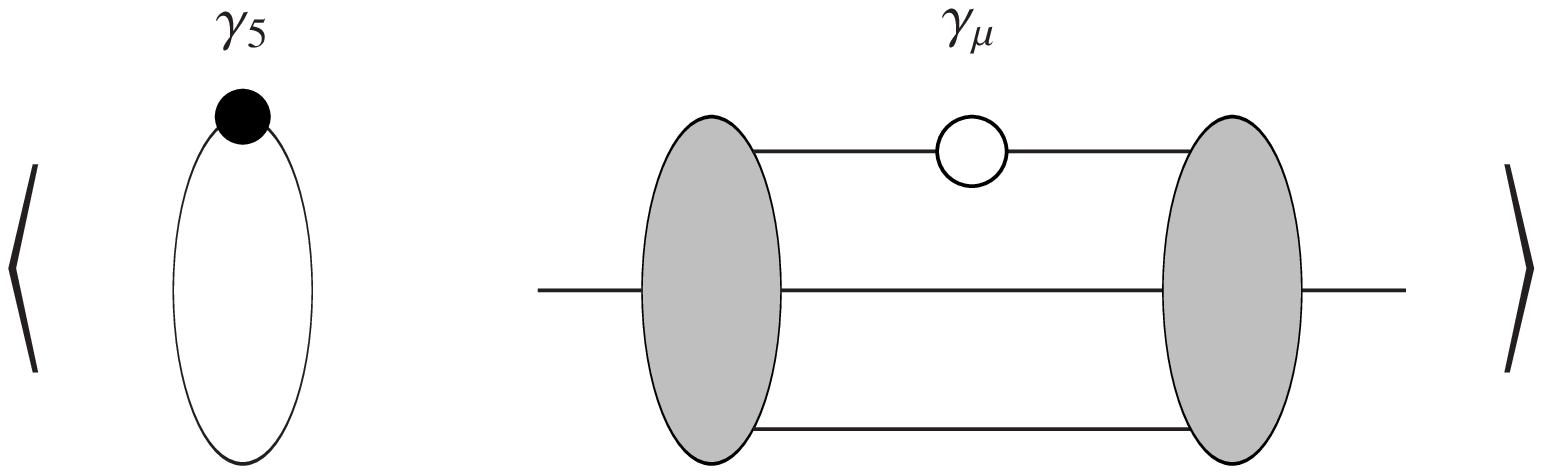,width=10cm,clip=}
\end{center}
\vspace*{-1cm}
\caption{Disconnected insertion of the pseudoscalar density to lowest order. Gluon lines are omitted.}
\label{fig2}
\end{figure}

The topological $\theta$ term (\ref{Stheta}) polarizes the vacuum. Diagrammatically it solely contributes to internal gluon lines. Similarly, the flavor-singlet pseudoscalar density in (\ref{Itheta}) and (\ref{Saction}) interacts with the nucleon through quark-line disconnected diagrams only~\cite{Aoki:1990ix,Guadagnoli:2002nm}. This is sketched in Fig.~\ref{fig2}. Consequently, the quark propagators in the nucleon matrix element (\ref{me}) are computed with the action $S_0^q$, neglecting the $S_\theta^q$ term.

We denote the two-point function of a nucleon of momentum $\vec{p}$ in the
$\theta$ vacuum by $G_{NN}^\theta(t,\vec{p})$. The phase factor $\alpha$
is obtained from the ratio of two-point functions
\begin{equation}
\begin{split}
{\rm Tr}\, [G_{NN}^\theta(t;0) \Gamma_4] & =  \frac{1+\cos{2\alpha(\theta)}}{2}\,\frac{1}{2}\, |Z_N|^2\, {\rm
  e}^{-m_N^\theta t}\,, \\
{\rm Tr}\, [G_{NN}^\theta(t;0) \Gamma_4 \gamma_5] & = i\, \frac{\sin{2\alpha(\theta)}}{2}\, \frac{1}{2}\, |Z_N|^2\, {\rm
  e}^{-m_N^\theta t}\,,
\end{split}
\label{def}
\end{equation}
where $\Gamma_4 = (1+\gamma_4)/2$. Equation (\ref{def}) defines $m_N^\theta$, the nucleon mass for the action (\ref{Saction}), and $Z_N$. The form factor $F_3(q^2)$ can be extracted from the ratio of three-point and two-point functions, generalizing the methods developed in~\cite{Capitani:1998ff}
\begin{equation}
\begin{split}
R_\mu(t^\prime,t;\vec{p}^{\,\prime},\vec{p}) &= \frac{G_{NJ_\mu
    N}^{\theta\, \Gamma}(t^\prime,t;\vec{p}^{\,\prime},\vec{p})}{{\rm Tr}\,
    [G_{NN}^\theta(t^\prime;\vec{p}^{\,\prime})\Gamma_4]} \\[0.5em]
    &\times \left\{\frac{{\rm
    Tr}\,[G_{NN}^\theta(t;\vec{p}^{\,\prime})\Gamma_4]\,  
    {\rm Tr}\,[G_{NN}^\theta(t^\prime;\vec{p}^{\,\prime})\Gamma_4]\,
    {\rm Tr}\,[G_{NN}^\theta(t^\prime-t;\vec{p})\Gamma_4]}
    {{\rm Tr}\,[G_{NN}^\theta(t;\vec{p})\Gamma_4]\, 
    {\rm Tr}\,[G_{NN}^\theta(t^\prime;\vec{p})\Gamma_4]\,
    {\rm
    Tr}\,[G_{NN}^\theta(t^\prime-t;\vec{p}^{\,\prime})\Gamma_4]}\right\}^{1/2}
    \\[0.75em] 
&= \sqrt{\frac{E^{\theta\,\prime} \,E^\theta}
   {(E^{\theta\,\prime}+m_N^\theta) \,
    (E^\theta+m_N^\theta)}} \, F(\Gamma,\mathcal{J}_\mu)\,,
\end{split}
\label{rw}
\end{equation}
where $G_{NJ_\mu N}^{\theta\, \Gamma}(t^\prime,t;\vec{p}^{\,\prime},\vec{p})$
is the three-point function, with $t^\prime$ being the time location of
the nucleon sink and $t$ the time location of the current insertion, and the
function $F(\Gamma,\mathcal{J}_\mu)$ is
\begin{equation}
%\begin{split}
F(\Gamma,\mathcal{J}_\mu) = \frac{1}{4}\, {\rm Tr}\, \Gamma %&
\left[{\rm
    e}^{i\alpha(\theta)\gamma_5} \frac{E^{\theta\,\prime}\gamma_4
    -i\vec{\gamma}\vec{p}^{\,\prime} + m_N^\theta}{E^{\theta\,\prime}}\,
    {\rm  e}^{i\alpha(\theta)\gamma_5}\right]\,%\\[0.5em]
%&\times 
\mathcal{J}_\mu\, \left[{\rm
    e}^{i\alpha(\theta)\gamma_5} \frac{E^{\theta}\gamma_4
    -i\vec{\gamma}\vec{p} + m_N^\theta}{E^{\theta}}\,
    {\rm  e}^{i\alpha(\theta)\gamma_5}\right]
%\end{split}
\end{equation}
with $\mathcal{J}_\mu$ given in (\ref{current}). The three-point functions are
calculated for various choices of nucleon polarization, $\Gamma = \Gamma_4$,
$i\Gamma_4\gamma_5\gamma_1$, $i\Gamma_4\gamma_5\gamma_2$ and $i\Gamma_4\gamma_5\gamma_3$. For $J_\mu$ we take the local vector current $\bar{q}\gamma_\mu q$.

\begin{figure}[t!]
\vspace*{-1.25cm}
\begin{center}
\epsfig{file=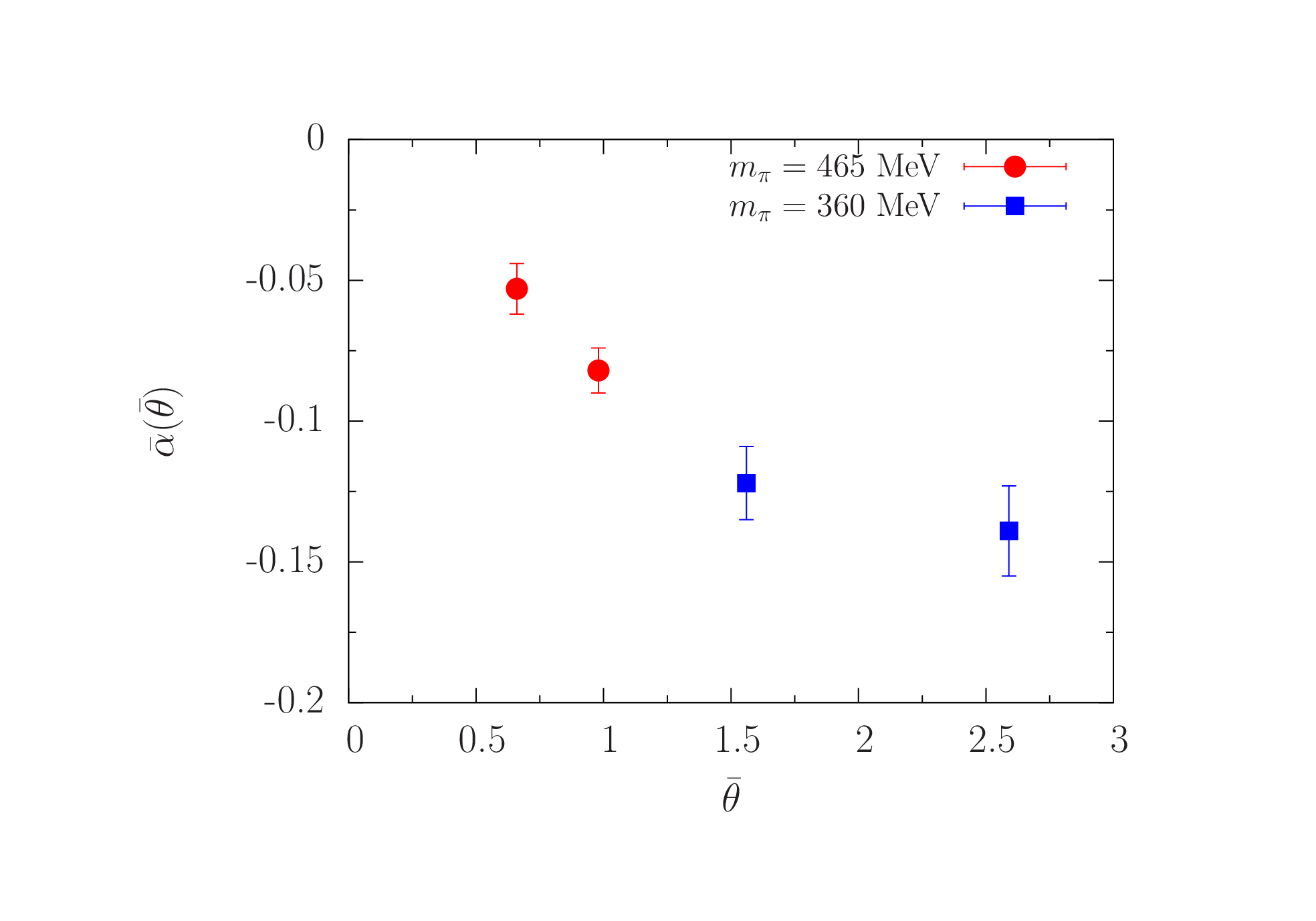,width=13cm,clip=}
\end{center}
\vspace*{-1cm}
\caption{The phase factor $\bar{\alpha}(\bar{\theta})$ as a function of $\bar{\theta}$ for our two sets of quark masses.}
\label{fig3}
\end{figure}

\section{Results}

In physical units, the pion and kaon masses are~\footnote{It is to be noted that the pseudoscalar mass at our flavor symmetric point are somewhat larger than the physical value $\displaystyle \sqrt{\left(m_{K^0}^2+m_{K^+}^2+m_{\pi^+}^2\right)/3}=413\,\mbox{MeV}$.}
\begin{equation}
\begin{tabular}{c|c|c|c}
$\kappa_\ell $ & $\kappa_s$ & $m_\pi\,\mbox [MeV]$ & $m_K\,\mbox [MeV]$ \\ \hline
0.12090 & 0.12090 & 465(13) & 465(13) \\
0.12104 & 0.12062 & 360(10) & 505(14)
\end{tabular}
\end{equation}
To a good approximation $2m_K^2+m_\pi^2 = \mbox{constant}$, in accord with the leading order chiral expansion $2m_K^2+m_\pi^2= 6\, B_0\, \bar{m}$.

%To make contact with phenomenology, the $\theta$ angle needs to be %renormalized. The result is
%\begin{equation}
%\theta^R=\frac{\hat{m}}{\hat{m}^R}\,\frac{1}{Z_P}\,\theta \,, \quad %\bar{\theta}^R=\frac{\hat{m}}{\hat{m}^R}\,\frac{1}{Z_P}\,\bar{\theta} \,.
%\end{equation}
%The quark masses renormalize as~\cite{Bietenholz:2011qq}
%\begin{equation}
%m_q^R=Z_m^{NS}\,\left(m_q+\alpha_Z\,\bar{m}\right)\,.
%\end{equation}
%A recent calculation~\cite{Constantinou:2014fka} gave %$Z_P\,Z_m^{NS}=Z_P/Z_S^{NS}=0.664(6)$. The parameter $\alpha_Z$ can be %estimated from the ratio of valence to sea quark %masses~\cite{Bietenholz:2011qq}. An updated value is $\alpha_Z=0.82(8)$.

\begin{figure}[t!]
\vspace*{-1.25cm}
\begin{center}
\epsfig{file=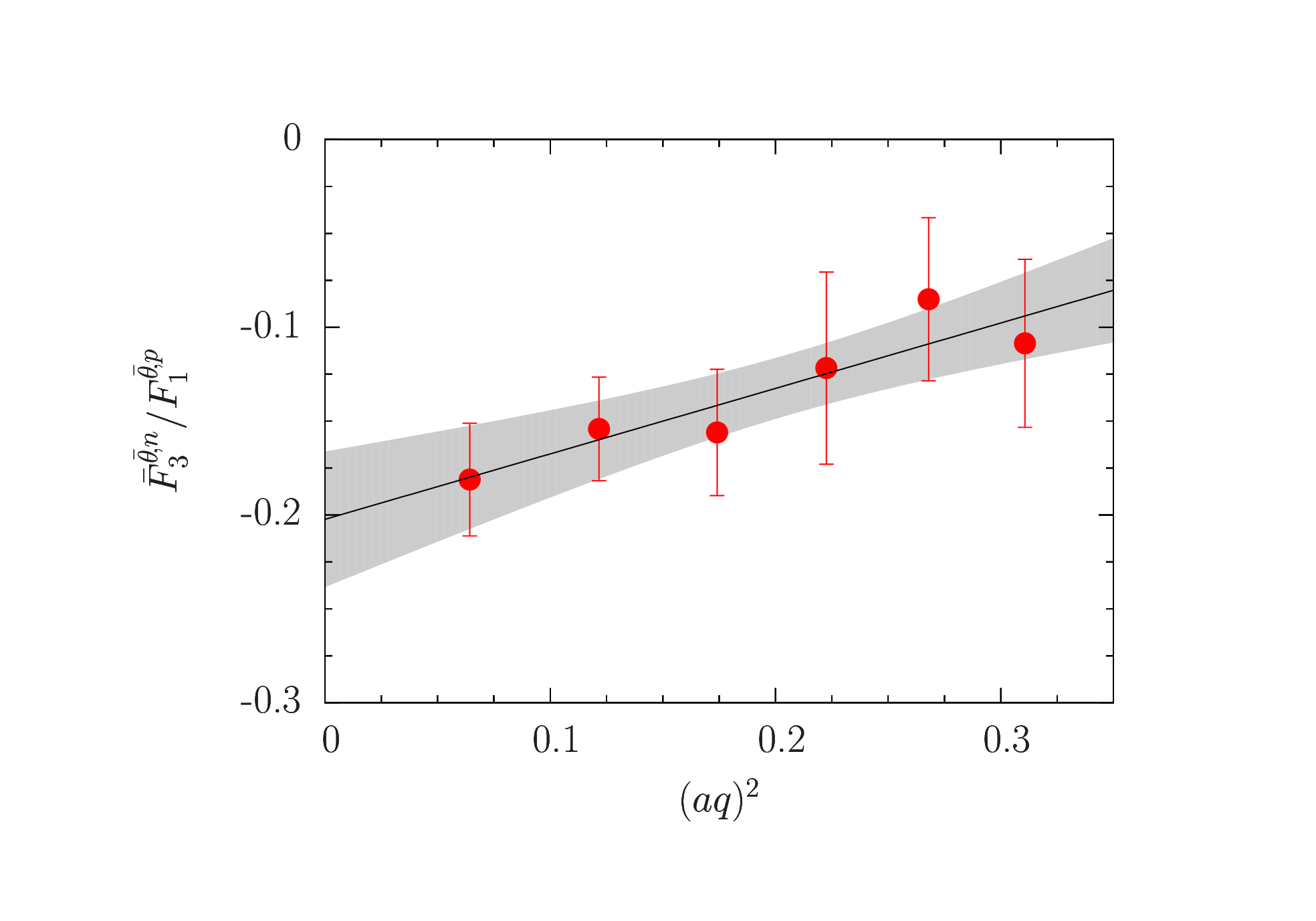,width=13cm,clip=}
\end{center}
\vspace*{-1cm}
\caption{The ratio of form factors $\bar{F}_3^{\bar{\theta},n}/F_1^{\bar{\theta},p}$ for $\kappa_\ell=\kappa_s=0.12090$ and $\lambda=0.005$.}
\label{fig4}
\end{figure}

At imaginary values of $\theta$, both $\alpha(\theta)$ and $F_3^\theta$ are imaginary. Thus, we can write 
\begin{equation}
\alpha(\theta) = i\, \bar{\alpha}(\bar{\theta}) \,, \quad F_3^\theta = i\, \bar{F}_3^{\bar{\theta}} \,.
\end{equation}
In Fig.~\ref{fig3} we show the results for the phase factor $\bar{\alpha}(\bar{\theta})$, and in Fig.~\ref{fig4} we show the form factor $\bar{F}_3^{\bar{\theta},n}$ 
\clearpage

\begin{figure}[t!]
\vspace*{-1.25cm}
\begin{center}
\epsfig{file=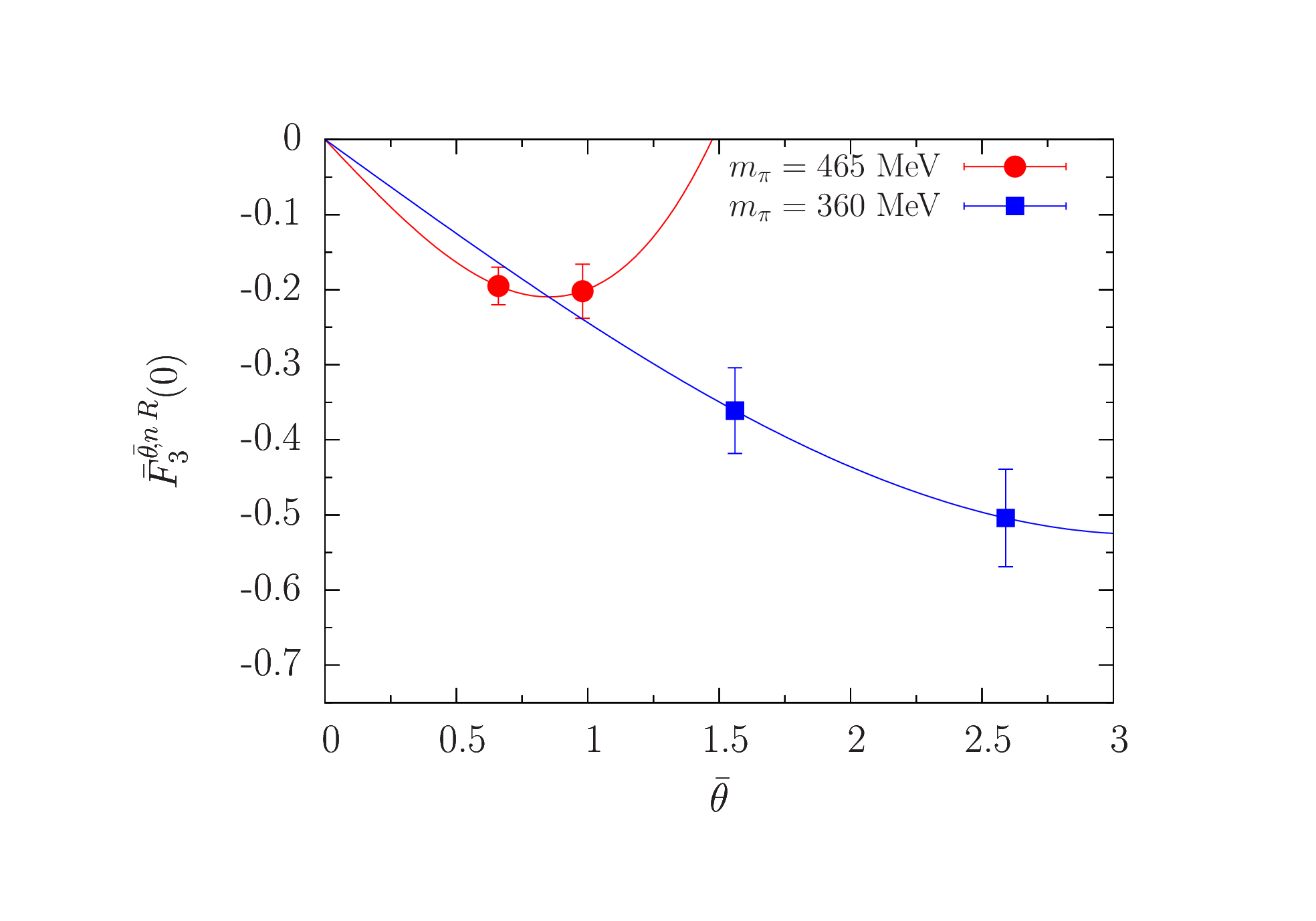,width=13cm,clip=}
\end{center}
\vspace*{-1cm}
\caption{The renormalized form factor $\bar{F}_3^{\bar{\theta},n\;R}(0)$ as a function of $\bar{\theta}$, together with a linear plus cubic extrapolation, $\bar{F}_3^{\bar{\theta},n\;R}(0) = A\,\bar{\theta} + B\,\bar{\theta}^3$, to $\bar{\theta}=0$.}
\label{fig5}
\end{figure}

\clearpage
\noindent
of the neutron divided by $F_1^{\bar{\theta},p}$ of the proton for ensemble 2. If the radii of the two form factors are close to one another, the $q^2$ dependence is largely cancelled out in the ratio. Indeed, the ratio shows only a mild $q^2$ dependence and thus may be extrapolated linearly to $q^2=0$. The extrapolated value is the renormalized form factor $\bar{F}_3^{\bar{\theta},n\;R}(0)$, using the fact that $\bar{F}_1^{\bar{\theta},p\;R}(0)=1$, from which we obtain the electric dipole moment (\ref{edm}). In Fig.~\ref{fig5} we show our results for $\bar{F}_3^{\bar{\theta},n\;R}(0)$ as a function of $\bar{\theta}$ for our two sets of quark masses. It should be noted that the actual expansion parameter is $\lambda$, given in (\ref{lambda}), which is a very small number. 

Ultimately, we are only interested in $\bar{F}_3^{\bar{\theta}}(0)$ (we drop the superscripts $\mbox{\small $n, R$}$ on $F_3$ from now on) at very small values of $\bar{\theta}$. Even so, we do not have sufficient data to constrain the extrapolation of $\bar{F}_3^{\bar{\theta}}(0)$ to $\bar{\theta}=0$. This will result in a systematic error. To estimate the error, we have employed a linear plus cubic fit, $A\, \bar{\theta} + B\, \bar{\theta}^3$, a Pad\'e fit, $A\, \bar{\theta}/(1 + B\, \bar{\theta}^2)$, allowing for corrections of $O(\bar{\theta}^5)$ and higher, as well as a linear fit, $A\, \bar{\theta}$, to the lowest $\bar{\theta}$ point each. We identify the central value of $A$ with the derivative of $\bar{F}_3^{\bar{\theta}}(0)$ at $\bar{\theta}=0$, $\bar{F}_3^{(1)}(0)$. The coefficient $A$ of the linear plus cubic fit shown in Fig.~\ref{fig5} turns out to be close to the central value. The error of $\bar{F}_3^{(1)}(0)$ is estimated to be the largest deviation of $A$ from the central value. 
\begin{figure}[t!]
\vspace*{-1.25cm}
\begin{center}
\epsfig{file=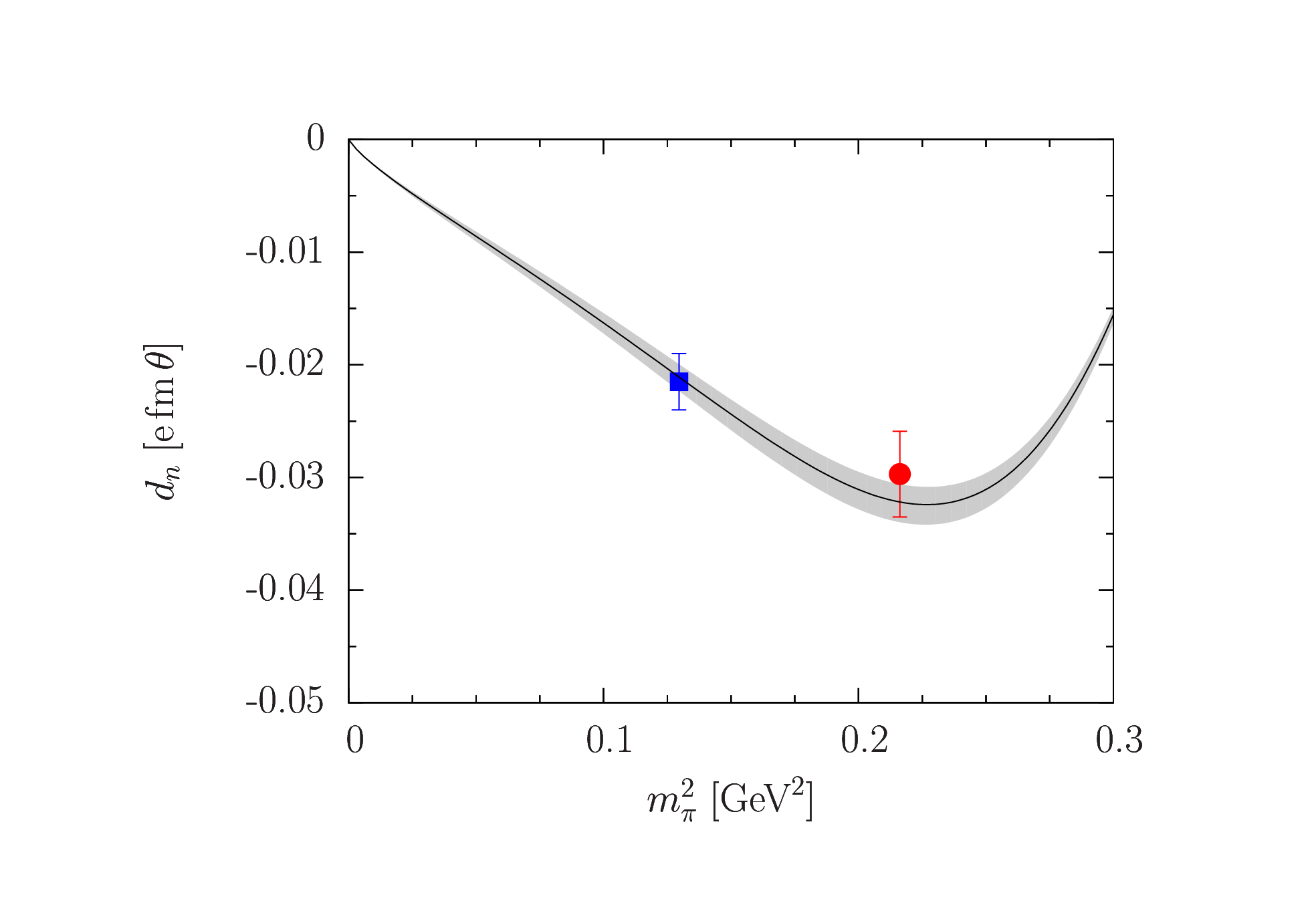,width=13cm,clip=}
\end{center}
\vspace*{-1cm}
\caption{The dipole moment of the neutron extrapolated to the physical point along the path $\bar{m}=\mbox{constant}$.}
\label{fig6}
\end{figure}
After continuing $\theta$ and $F_3^\theta(0)$ back to real values, we finally obtain, writing $d_n=e\,F_3^{(1)}(0)\,\theta/2m_N$, 
\begin{equation}
\begin{tabular}{c|c|c}
$m_\pi\,[\mbox{MeV}]$ & $m_K\,[\mbox{MeV}]$ & $d_n\,[\mbox{e\,fm\,$\theta$}]$ \\ \hline
465(13) & 465(13) & $-0.0297(38)$\\
360(10) & 505(14) & $-0.0215(25)$
\end{tabular}
\label{dnres}
\end{equation}
To extrapolate (\ref{dnres}) to the physical point, we make use of the analytic expressions derived from covariant $U(3)_L\times U(3)_R$ baryon chiral perturbation theory in~\cite{Guo:2012vf} to NLO, with the additional constraint $2m_K^2+m_\pi^2=\mbox{constant} \propto \bar{m}$. This basically involves one free low-energy constant, $w_a(\mu)$, only. A fit to the lattice data gives $w_a(\mu=1\,\mbox{GeV})=0.04(1)\,\mbox{GeV}^{-1}$. The result of the fit is shown in Fig.~\ref{fig6}. Note that $d_n$ vanishes at $2m_K^2-m_\pi^2=0$ due to the constraint $\bar{m}=\mbox{constant}$. At the physical point this finally leads to
\begin{equation}
d_n = - 0.0039(2)(9)\, [e\,\mbox{fm}\,\theta] \,.
\label{final}
\end{equation}
The first error is purely statistical. The second error is a conservative estimate of NNLO effects. It covers the naive result from a polynomial extrapolation, $d_n = - 0.0043\, [e\,\mbox{fm}\,\theta]$.

%If the electric dipole moment is known experimentally, 
Our result (\ref{final}) translates into constraints on CP violating contributions to the action at the quark and gluon level. The current experimental bound on the electric dipole moment of the neutron is~\cite{Baker} 
$\displaystyle |d_N^n| \leq 2.9 \,\times \, 10^{-13} \, [e\,\mbox{fm}]$. 
Combining this bound with (\ref{final}), we arrive at the upper bound on $\theta$,
\begin{equation}
|\theta| \lesssim 7.4 \,\times 10^{-11} \,.
\end{equation}
%A recent calculation~\cite{Guo:2012vf} of the electric dipole moment of the %neutron from $U(3)_L\times U(3)_R$ chiral perturbation theory with input from %the lattice~\cite{Shintani:2008nt,Shintani:2012uba} gives $d_n = - %0.0029(9)\, [e\,\mbox{fm}\,\theta]$. This translates into the upper bound %$|\theta| < 1.0 \,\times 10^{-10}$. 

\section{Conclusions}

It should be noted that in this exploratory work we have not included contributions from disconnected insertions of the electromagnetic current. However, since these contributions vanish exactly at the flavor symmetric point, we do not expect them to have a significant effect to our conclusions. It remains to be seen how big they are at the physical point. 

The vacuum angle $\theta$ renormalizes as $\theta^R = (Z_S^S/Z_P)\,\theta$, where $Z_S^S$ and $Z_P$ are the renormalization constants of the flavor-singlet scalar density and the pseudoscalar density, respectively. In the continuum $Z_S^S/Z_P=1$. A caveat of our calculations is that clover fermions, though $O(a)$ improved, break chiral symmetry at finite lattice spacings. On our present lattices $Z_S^S/Z_P = 0.8 - 0.9$~\cite{Constantinou:2014fka,Bietenholz:2011qq,Chambers:2014pea}, which might imply a systematic error of $O(10\%)$. 

To sum up, we have successfully computed the electric dipole moment of the neutron $d_n$ from simulations of $2+1$ flavor lattice QCD at imaginary vacuum angle $\theta$, using the axial anomaly to rotate the topological charge density into a flavor singlet pseudoscalar density in the fermionic action. Only disconnected insertions of the pseudoscalar density contribute to the dipole moment, which required the generation of new gauge field ensembles with the modified action (\ref{Saction}). Clearly, our results will have to be substantiated by simulations on larger lattices, at smaller pion masses and  smaller lattice spacings, as well as for a wider range of $\lambda$ parameters. This is a challenging task, which we hope to report on in due course. 

%This study paves the way for future simulations on larger lattices and at %smaller quark masses. 
 
\section*{Acknowledgements}

This work has been partly supported by DFG, Grant Schi 422/9-1, the Australian Research Council, Grants FT100100005 and DP140103067, DFG and NSFC through the Sino-German CRC 110, and NSFC, Grant 11165005. The numerical calculations were carried out on the BlueGeneQ at FZ J\"ulich, Germany and on the BlueGeneQ at EPCC Edinburgh, UK using DIRAC 2 resources.

\end{document}